\documentstyle[multicol,prl,aps,epsf,psfig,epsfig]{revtex}
\begin{document}

\title
{ Complexities of social networks: A Physicist's perspective}
\author
{
  Parongama Sen
}
\address
{
 Department of Physics, University of Calcutta,
    92 Acharya Prafulla Chandra Road, Kolkata 700009, India.
}

\maketitle
Preprint no:CU-Physics 02-2006

\begin{multicols}{2}                                                                                

\section{Introduction}
A large number of natural phenomena in the universe go on oblivious of the
presence of any living being, let alone a society. 
Fundamental science subjects like  physics and chemistry deal with laws
which  would remain unaltered in the absence of life.
In biological science, which deals with
living beings, the idea of a society may exist in a basic  form, 
e.g.,  as food chains, struggle for existence etc.,  among different species 
whereas present day human society has many more layers as factors like 
politics, economics or psychology dictate its evolution to a large extent.   

A  society of human beings can be conceived in different ways as it involves  
different kinds of contacts.  Based on a certain kind of interaction, 
a collection of human beings may be thought of as a network where the
individuals are the nodes and the links are formed whenever two
of them interact in the defined way \cite {wasser}. 
An interesting aspect of several such social networks is that 
these show small  world effect, a phenomenon which has been shown
to exist in diverse kinds of networks \cite{watts}. 
Small world effect and other 
interesting  features shared by  real world networks of different nature 
have triggered off a tremendous activity of research among scientists of 
different disciplines \cite{ba_review,ba_book,dorobook}.
Physicists' interests in networks  lie in the fact that these
show interesting phase transitions as far as equilibrium 
and dynamic properties are concerned. The tools of statistical
physics come handy in the analytical and numerical studies  of  networks. 
Also, physics
of systems embedded in small world networks raise interesting questions.

In this article we discuss the properties of social networks which
have received much attention during  recent years. 
The topic of social networks is vast and covers many aspects. 
Obviously, all the issues
are too difficult to address  in a single review 
and thus we have tried to give an overview of the subject,
emphasising a physicist's perspective whenever possible.

\section{The beginning: Milgram's experiments}

 The first
real world network which showed the small world effect
  was a social network. This was the result of  some 
experimental studies  by the social 
psychologist, Milgram \cite{milgram}. In Milgram's experiments, 
 various recruits
in remote places of  Kansas and Nebraska were asked to forward  letters to 
specific addresses in Cambridge and Boston in Massachusetts.  
A letter had to be hand-delivered 
only through persons known on first-name basis. 
Surprisingly, it was found that on an average six people were required for a
successful 
delivery. The number 6 is approximately $\log(N)$, where $N$ is the total
number of people in the USA. 
In this particular experiment, if a person A gave the letter to B,  
A and B are said to have a link between them. If B next hands over 
the mail to C, the effective ``distance'' between A and C is 2; while
between A and B, as well as between B and C, it is 1. 
Defining distance in this way, Milgram's experiments show that two persons in a society are 
separated by an average distance of six steps.
This property of having a small average distance 
in the society is what is   known as the 
small world effect. 

It took around thirty years to realise that this small world effect 
is not unique to the human society but rather possessed by
a variety of other real  (both natural and artificial) networks.  
These networks  include social networks, Internet and
WWW network, power grid network, biological networks,
  transport networks etc.

\section{Topological properties of networks}

Before  analysing  a society from the network point of view, 
it will be useful to
summarise the topological properties characterising  
common small world networks.

\begin{figure}[b]
\noindent \includegraphics[clip,width= 4cm]{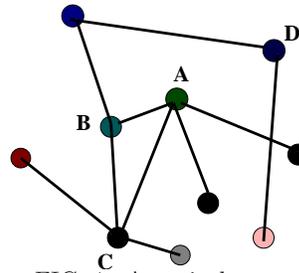}
\caption{A typical network. The shortest distance from A to C is 1 while 
that from
A to D is 3. A,B,C form a cluster.}
\label{fig1}
\end{figure}

 A network is nothing
but a graph having nodes as the vertices and links between nodes as edges.
A typical network is shown in Fig. \ref{fig1}  where there are ten nodes and ten 
links.

The shortest distance $S$ between any two nodes A and B 
is the number of edges on the  shortest path  from A to B 
through connected nodes. In Fig. \ref{fig1}, the shortest path from A to D 
goes through three edges and $S_{AD}=3$.  
The diameter $D$ of a network   is the largest of 
the  shortest distances $S$  
and in a small world network (SWN), both the average shortest distance 
$\langle S\rangle$ and $D$ scale in the same way with $N$, the number of nodes in the network.

Erd\"os and R\'enyi studied the random graph \cite{erdos} in which any two 
vertices have a finite probability to get linked.  In this network or graph, 
both $D$ and $\langle S\rangle$  were found to vary as $\log(N)$. 
 (This result is true when a minimum number of edges is present in 
the graph so that a giant structure is formed; if in the graph with $N$ nodes, 
the connectivity probability of any two nodes is $p$, a giant structure is formed for $pN > 1$.)

The network of human population, or for that matter, many other
networks can hardly be imagined to be  a random network, although
the latter has the property of a small average  shortest distance between nodes. 
The other factor which is bound to be present in a social network (like
the one considered by Milgram) is the clustering tendency. This precisely 
means that if 
A is linked to both B and C there is a strong likelihood that C is also 
linked to B.
This property is not expected to be present in a random network.

Let $a_{ij}$ define the adjacency matrix of the network; $a_{ij}=1 $ if
nodes $i$ and $j$ are connected and zero otherwise. 
Then one can quantify the clustering 
coefficient $C_i$ of node $i$ as

\begin{equation}
C_i= \Sigma_{j_1\neq j_2} \frac 2{k_i(k_i+1)}{a_{j_1j_2}},
\end{equation}
where $j_1,j_2$ are nodes connected to $i$ and $k_i=\Sigma_j a_{ij}$, the
total number of links possessed by node $i$, also known as its degree
(for example in Fig. \ref{fig1}, the degree of node A is 4).
Measuring the average clustering coefficient $C=\frac{1}{N}\Sigma_i C_i$ 
in several networks, it was 
observed  that the clustering coefficient was order of magnitude higher 
than that of a  random network with the same number of vertices and edges. 
Thus  it was concluded that the real-world networks were quite different from
a random network as far as clustering property is concerned.
Networks with a small value of $\langle S \rangle$ or $D$ ($O(\log(N))$)
together with a  clustering coefficient much larger than the 
corresponding random network were given the name small world network.
Mathematically it is possible to distinguish between 
a small world network and a random network by comparing their clustering 
coefficients.

Another type of network may be conceived in which the clustering property is
high but the average shortest distance between nodes is comparable to $N$. Such networks
are called regular networks and thus the small world networks lie in between 
regular and random networks.

One can find out the probability distribution of the  number of neighbours
of a node, commonly called the degree, of a network.
An interesting  feature  revealed in many real world networks is the
scale-free property \cite{bascience}. This means that the degree distribution $P(k)$ (i.e.,
the probability that a node has degree $k$) shows the
behaviour $P(k) \sim k^{-\gamma}$, 
 implying the presence of  a few nodes which are very highly
connected. These highly connected nodes are called the hubs of the network.

It may be mentioned here that in the random graph, $P(k)$ has a different
behaviour,
\begin{equation}
P(k) = e^{-\langle k\rangle}\frac {{\langle k\rangle}^k}{k!},
\end{equation}
where $\langle k\rangle$ is the average degree of the network.
$P(k)$ therefore  follows a Poisson distribution here decaying rapidly with $k$.

The above  features constitute the main   properties of networks.
Apart from these, many other characteristics have been detected 
and analysed as research in small world networks increased by leaps and bounds.
Some of these, e.g., closeness centrality or betweenness 
centrality  were  already  quite familiar to social scientists \cite{wasser}.\\
{\it{Closeness centrality:}} The measure of the average shortest distance
of a node to the other nodes in a network is its closeness centrality.\\ 
{\it Betweenness centrality}: The fraction  of shortest paths passing through 
a node is 
its betweenness \cite{freeman,goh2}. The more is the betweenness of a node,
the more important it is in the network since its absence will affect the
small world property  to a great extent. 
It is not necessarily true that nodes with maximum degree
will have the largest closeness or betweenness centrality.\\
{\it Remaining degree distribution:}
If we arrive at a vertex  following a random edge,
the probability that it has degree $k$ is $kP(k)$. The remaining degree distribution $q_k$, which is the probability that the node has $k$ other edges, is given by
\begin{equation}
\label{remain}
q_k = \frac{(k+1)P({k+1})}{\Sigma_j jP(j)}.
\end{equation}\\
{\it Assortativity} : This measure is for the correlation between 
degrees of nodes 
which share a common
edge. 
A straightforward measure will be to calculate the average degree
$\langle k_{nn} \rangle $ of the neighbours of a vertex with degree $k$.
If $\frac {d\langle k_{nn} \rangle }{dk} > 0$, it  will mean a positive
correlation or assortativity in the  network. A negative value of the 
derivative
denotes disassortativity and  a zero value would mean no correlation.
A more rigorous method of calculating the assortativity is 
given in \cite{newmanprlasso}, 
where  one defines a
quantity $r$ as 
\begin{equation}
\label{assort}
r = \frac{M^{-1}\Sigma_i j_ik_i - [M^{-1}\Sigma_i\frac{1}{2}(j_i+k_i)]^2}
{M^{-1}\Sigma_i \frac{1}{2}(j_i^2 + k_i^2) - [M^{-1}\Sigma_i\frac{1}{2}(j_i+k_i)]^2},
\end{equation}
where $j_i$ and $k_i$ are the degrees of the vertices connected by the $i$th 
edge ($i=1,2,....,M$) and $M$ is the total number of edges in the network.
Again, high assortativity means that two nodes which are both highly connected tend
to be linked and $r > 0$. A negative value of $r$ implies that nodes with 
dissimilar degrees are more likely to get connected. A zero value implies no
correlation of node degrees and their connectivity.

{\it {Community structure}}:
This is a property which is highly important for the social networks.
More often than not we find a society divided into  a number of communities, 
e.g, based on profession, hobby, religion  etc.  For the scientific 
collaboration network, communities may be formed 
with scientists belonging to different fields of research, for example,  physicists,
mathematicians or biologists. Within a community also, there may be different divisions
like physicists may be classified into different groups, e.g.,  high energy physicists, 
condensed matter physicists and so on.

It is clear that the properties of a network simply depends on the
way the links are distributed among the vertices, or to be precise, 
on the adjacency matrix. Till now we have not specified anything about the 
links.
Links in the networks may be both directed and undirected, e.g., in a 
e-mail network \cite{ebel-email}, if A sends a mail to B, we have a directed link from A to B. 
Edges may also be weighted; weights may be defined in several ways 
depending on the type of network. In the weighted collaboration network,
 two authors sharing a large number of publication have  
a link which has more weight than that between two authors  who have collaborated 
fewer times.

\section{Some prototypes of small world networks}

At this juncture, it is useful to describe a few important  prototype small world 
network models which have been considered to mimic the properties of 
real networks. 

\subsection{Watts and Strogatz (WS) network}

This was the first network model which was successful in reproducing the
features of small diameter and   large clustering coefficient of a
network. The small world effect in    networks of varied nature indicated 
a similarity in the  underlying structure of the networks.
Watts and Strogtaz \cite{watts} conjectured  that the geometry of 
the networks have some common features responsible for the
small world effect.
In their model, the nodes are placed on a 
ring. Each node has connection to $k$ number of nearest neighbours initially.
With probability $p$, a link is then rewired to form a random long ranged link.

At $p=0$, the shortest paths scale as $N$ and the clustering coefficient of the
network is quite high as it behaves as a regular network with
a considerable  number of nearest neighbours. 
The remarkable result was, even with $p\to 0$, the diameter of the 
network is small ($O(\log(N))$).
The clustering coefficient on the other hand remains high even when $p\neq 0$ 
unless $p$ approaches unity. 
Thus for $p\to 0$, the network has a small diameter as well as high clustering coefficient, i.e., it is a small world network. 
For $p \to 1$, the network ceases to have a large clustering coefficient and
behaves as a random network. This  model thus 
displays phase transitions from a regular to a  small world to a  random graph
 by varying a single parameter $p$. 
Later  it was shown to have mean field behaviour in the small world phase \cite{expo}.

The degree distribution $P(k)$ in this network, however, did not have a power
law behaviour  but showed  an exponential decay with a peak at $\langle k\rangle $.

\subsection{Networks with small world and scale-free property}\index{heading}

Although the WS  model was successful in showing the small
world effect, it did not  have a power law degree 
distribution.
The discovery of scale-free property in many real
world networks required the  construction of  a model which would 
have small world as well as scale-free property.

Barab\'asi and Albert (BA) \cite{bascience} proposed an evolving model in which one starts with
a few nodes linked with each other. Nodes are then added one by one. An incoming node will have a probability $\Pi_i$ to get attached to the $i$th node already 
existing in the network 
according to the rule of preferential attachment which means that 
\begin{equation}
\Pi_i = k_i/\Sigma k_i,
\end{equation}
where $k_i$ is the degree of the $i$th node.
This implies that a node with higher degree will get more links as
the network grows such that it has a ``rich gets richer'' effect.
The results showed a power law degree distribution with 
exponent $\gamma=3$.
While the average shortest distance grows with $N$    slower than $\log(N)$ 
in this 
network,  
the clustering coefficient vanishes in the 
thermodynamic limit.
Several other
network  models have been conceived later as variants of the BA network 
which allow a finite value of the clustering coefficient.
Also, scale-free networks have been achieved using algorithms other than
the preferential  attachment rule \cite{HA} or even without considering
a growing network \cite{geo}.

\subsection{Euclidean and time dependent networks}

In many real world networks  the nodes are embedded
on a Euclidean space 
and the link length distribution shows a strong 
distance  dependence \cite{yook,katz,nagpaul,geo2,gastner,olson,anjan}. 
Models of Euclidean networks have been constructed for both static and growing
networks \cite{waxman,klein,blumen,psbkc,mouk,ps_presi,psmanna1,psmanna2,psmanna3}. In static models, transition between regular, random and 
small world phases may be obtained by manipulating  a single parameter occurring in the
link length distribution
\cite{blumen,psbkc,mouk,ps_presi}. In a growing model,
a distance dependent factor is incorporated in a generalised 
preferential attachment 
scheme \cite{psmanna1,psmanna2,psmanna3} giving rise to  a  
transition between scale-free and non scale-free network.

Aging  is also another factor which is present in many evolving networks, e.g.,
the citation network \cite{ps_hajra1,redner1}. 
Here the time factor plays an important role in the linking scheme.
Aging of nodes has been  taken into account 
in a few theoretical models 
where the aging factor is  incorporated in the attachment probability suitably. 
Again one can achieve a transition from scale-free to 
non scale-free network by appropriately tuning the parameters
\cite{doro,hua,ps_hajra2}.

\section{Social networks: classification and examples}

Social networks can be broadly divided into three classes.
The  social  networks in which  links are formed directly between
the nodes may be said to  belong to class A.
The friendship network is perhaps the most fundamental social network
of class A.
Other social networks like the networks of 
office colleagues, e-mail senders and receivers, sexual partners etc.
also belong to this class.

The second  class  of social networks consist of the  various
collaboration  networks which
are formed from bipartite networks. For example, in the movie actors'
network, two actors are said to have a link between them if they have
acted in the same movie. Similarly in the research collaboration
networks, two authors form a link if they feature in the same
paper. We classify these  networks as  class B social networks.
The difference between class A and class B networks is that in class A,
it is ensured that  two persons sharing a link have interacted at a personal
level while it is possible that in a collaboration act, two collaborators
sharing a link hardly know each other.

In some other social networks, which we classify as class C networks,
  the nodes are 
not human beings but the links connect people indirectly. Examples are citation networks and transport networks.
In a citation network, links are formed between two papers when one cites the
other.  Transport networks consist of railways, roadways and 
airways networks; the nodes here are usually cities which are linked by
air, road or rail routes. 


Real world data for class A networks, e.g., friendship,  acquaintance or sexual network is difficult to get
and usually available for small populations \cite{ba_review,mixing,tribe}.
Communication networks like e-mail \cite{ebel-email} and telephone \cite{telephone} networks
show small world effect  and power law degree distribution \cite{ba_review}.
Datasets  may be constructed artificially also, 
 e.g., as in \cite{WIW}, which 
shows small world effect in a friendship network.

Class B  networks have been studied extensively 
as many databases are available. These, in general, also show the small
world effect. 
In the movie actors network with over $2 \times 10^5$ nodes,
the average shortest path length is  3.65. Typically in co-authorship networks,
it varies between 2 to 10.
The degree distribution in these networks can be of different types
depending on the particular database.
Some collaboration networks have  a power law degree distribution, while   
there  exist  two different power laws regimes or 
an exponential cutoff in the degree distribution in other databases
 \cite{newman_coll,bara_coll}.

In Euclidean social networks like the 
scientific collaboration networks, links also
have  strong distance dependence \cite{katz,nagpaul,geo2,olson,anjan}, usually
showing a proximity bias.
Recently, in a collaboration network of authors of Physical Review Letters,
the study of time evolution of the link length  
distribution   was made by Sen et al \cite{anjan}.
It indicated   the
intriguing possibility  that in a few decades hence, scientists located 
at any distance may collaborate with equal probability thanks to the
communication revolution.

The citation network belonging to class C is also quite well studied.  
 It is an example of a  directed network. 
Analysis shows that the citation networks 
have very interesting behaviour of degree distributions and aging effects 
\cite{redner2,vazquez,redner1}.
While the out-degree (number of papers referred to) shows a Poisson-like distribution the in-degree (number of citations) 
has scale-free distribution. 

The aging phenomenon is particularly 
important  in citation networks; older papers are less cited, 
barring exceptions.
 Studies over a long time show \cite{redner1} that the number of cited 
papers of age $t$ (in years)
decreases exponentially with $t$, while the probability that a 
paper  gets cited after time $t$ of its publication,  decreases as a 
power law over an initial time period 
of typically 15-20 years and exponentially for large values of $t$.
These features can be reproduced  using a growing network model \cite{ps_hajra2} where the 
preferential attachment scheme contains an age-dependent 
factor such that an incoming node $j$ links with node $i$ with the probability 
\begin{equation}
\Pi_{ij}(\tau_i , k_i) = \frac {k_i^\beta \exp(-\alpha \tau_i)} {\Sigma_i k_i^\beta \exp(-\alpha \tau_i)},
\end{equation}
where $k_i$ and $\tau_i$ are the degree and age of node $i$ at that moment.

Although transport networks, also belonging to class C, do not involve  human beings directly as nodes,
they can have great impact on social networks of both class A and B types.
The idea that in a railway network, two stations are linked if at least one  
train 
stops at both, was introduced by Sen, Dasgupta et al
\cite{train} in a study of the Indian railway network. Some details of that
study is provided in the appendix. 

\section{Distinctive features of Social networks}

That  social networks of all classes have a small diameter is demonstrated 
in almost all  
real examples. Many (although not all) social networks  also show the scale-free property,
and the exponent for the degree distribution generally lies between 
1 and 3.

Social networks are characterised by three features: very high values of the clustering
coefficient, positive assortativity and community structure.

High clustering tendency is quite understandable - naively speaking, 
a friend of a friend is quite often  a friend also.
Again, in a collaboration network of scientists, whenever a paper is written by three
or more authors,  there is a large contribution to the clustering coefficient.

It is customary to   compare the clustering coefficients of a real network
to that of its corresponding random network to show that the real network 
has a much larger clustering coefficient. Instead of a totally  random model,
one could also consider a ``null'' model which is a 
network with the same number of nodes and edges,  having  also the same degree
distribution,  but otherwise random. The clustering
coefficients in non-social networks turn out to be comparable to that
of the null model.
In contrast, for social networks, this is not true.
                                                                                                                             
Let us consider the random  model.
Suppose  two neighbours of a vertex in this model have remaining  degrees
$j$ and $k$. Then there will be a contribution to the clustering if
these two nodes share an edge. With $M$ edges in the  network,
the number of edges shared by these two nodes is $jk/2M$. Both $j$ and $k$
are distributed according to eq. (\ref{remain}), and therefore the clustering coefficient is  
\begin{equation}
C = \frac {1}{2M} [\Sigma_k kq_k]^2 = {\frac {1}{N}}
\frac{[\langle k^2\rangle -\langle k\rangle^2]}{\langle k\rangle^3}.
\end{equation}
For networks not large enough this will still give a finite
value. 
One can compare it with the small network of the food web of organisms
in Little rock lake which has $N=72, \langle k\rangle = 21.0$ and $ \langle k^2\rangle=655.2$, giving $C=0.47$ which compares
well with
the actual value 0.40.
 It can be shown that for the null model with degree distribution
$P(k) \sim k^{-\gamma}$, the clustering coefficient remains
finite even for large $N$ 
when $\gamma < 7/3$.
The theoretical value of the clustering coefficient here is 
\begin{equation}
C \propto N ^{(7-3\gamma)/(\gamma-1)}.
\end{equation}
This is assuming that there can be more than one edge  shared by two vertices.
Ignoring multiple edges,
the clustering coefficient turns out to be \cite{newman_park} 
\begin{equation}
C = \Sigma_{jk} q_jq_k (1-e^{-jk/2M}).
\end{equation}
Comparison with non-social networks like the Internet, food webs, world wide webs
shows that  the difference between the values obtained theoretically
and the actual ones are minimum. For social networks however, the 
theoretical values are at least one order of magnitude
smaller than the observed ones.

Assortativity in social networks is  generally positive in contrast to non-social networks. 
For non-social networks like the Internet, WWW, biological networks etc., $r$ (eq. \ref{assort}) 
lies between -0.3 to -0.1 while for social networks 
like the scientific collaboration networks or company directors, $r$ is between 
0.2 to 0.4 \cite{newmanprlasso}.

It has been  argued that 
the large clustering coefficient and  the positive assortativity in social 
networks arise from the community structure.
Since the discovery of community structure in social networks, there
has been tremendous activity in this field and we devote the next section to
the details of these studies.

\section {Community Structure in social networks}

A society is usually divided into many groups and again the groups may 
regroup to form a bigger group. 
The community structure may reflect the self organisation of a network
to optimise some task performance, for example, in searching, optimal
communication pathways or even maximisation of productivity in collaborations.
There is no unique definition of a community but the general idea 
is  that the members within a community have more connections within 
themselves
and lesser with members belonging to other communities.

For small networks, it is possible to visually detect the community structure. But with the availability of large scale data in many fields in recent times it is essential to have good algorithms to detect the community structure.

\subsection{Detecting communities: basic methods}

A short review of the community detection methods is available in \cite{newman_review}, many other methods have developed  since then. Here we attempt to 
present a gist of some   basic  and some recently 
proposed  methods.

\subsubsection{Agglomerative and Divisive methods } 

The traditional method for detecting communities in networks is 
agglomerative hierarchical clustering \cite{jain}. In this method, each node is assigned it own cluster 
so that with $N$ nodes, one has $N$ clusters in the beginning.
Then the most similar or closest nodes are merged into a single cluster
following a certain prescription so that one is left with one cluster less. This can be continued 
till one is left with a single cluster. 

\begin{figure}
\noindent \includegraphics[clip,width= 8cm]{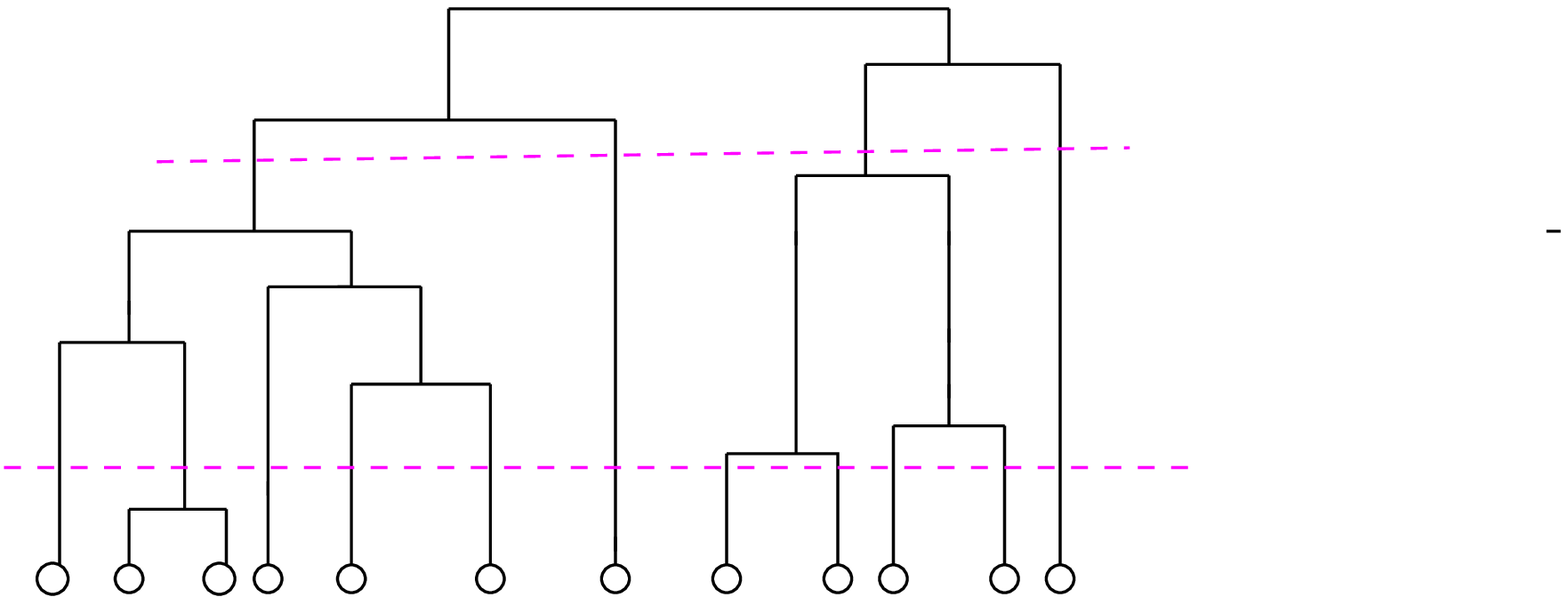}
\caption{ A typical dendogram. In the divisive algorithm, one explores from
top to bottom while for a agglomerative method, one goes from the bottom to the
top. The number of communities  is given by 
the number of intersection points of the horizontal dashed lines 
with the dendogram at different levels. The circles represent the nodes.}
\label{fig2}
\end{figure}

In the divisive method,  the nodes belong to a single community to begin with. By some
appropriate scheme,  edges are removed resulting in a  splitting of the 
communities into subnetworks in steps and ultimately all the nodes are
split into separate communities. Both the agglomerative and divisive methods give rise to what is known as 
a dendogram (see Fig. \ref{fig2}) which is a diagrammatic representation of the
nodes and communities at different times. While in the agglomerative method one 
goes from bottom to top, in the divisive method it is just the opposite. 

Girvan and Newman (GN) \cite{GN} proposed a divisive algorithm in which
edges are removed 
 using the measure of betweenness centrality (BC). 
 Generalising the idea of BC of nodes to BC of  edges, edge betweenness can be defined as 
the number of shortest paths between pairs of vertices that run along it.
An edge connecting members belonging to different
communities is bound to have a large betweenness centrality measure.
Hence in this method one calculates the edge betweenness at every step 
and removes the edge with the maximum measure. The process is continued till
no edge is left in the network. After every  removal of   an edge, the betweenness centrality is to be recalculated 
in this algorithm.

\subsubsection{A measure of the community structure identification}

Applying either of the above algorithms it may be 
possible to divide the network into communities. Even
the random graph with no community structure may 
be separated  into many classes. One needs to have a measure to see how good the 
obtained structure is. Also, the dendogram represents 
communities at all levels from a single to as many as 
the number of nodes and one should have  an idea  at which level the 
community division has been most sensible.

In case the network has a fixed well known community structure, 
a quantity to measure the performance of the algorithm
for the community detection may be the fraction of correctly classified nodes.

The other measure, known as the modularity of the network, can be
used when there is no ad hoc knowledge about the community structure
of the network. This is defined in the following way \cite{NG}. 
Let there
be $k$ communities in a particular
network. Then $e_{ij}$ is the fraction of edges which exist between nodes belonging to communities $i$ and $j$; clearly
it is a $k \times k$ matrix. 
The trace $Tr(e) = \Sigma e_{ii}$ then gives the fraction 
of all edges in the network that connect vertices in the same community and a 
good division into communities should give a high value of the trace.
However, simply the trace is not a good measure of
the quality of the division as placing all nodes in a single community 
will give trivially $Tr(e) =1$.
Let us now take the case when the network does not have a community structure; in 
this case links will be distributed randomly.
Defining the row (or column) sums $\alpha_i = \Sigma_j  e_{ij}$, one
gets the fraction of edges that connect to vertices in community $i$.
The expected number of fraction of links within a partition
is then simply the probability that a link begins at $i$, $\alpha_i$,
multiplied by the fraction of links that end at a node in $i, \alpha_i$.
So the expected number of intra-community links is just
$\alpha_i \alpha_i$.  
A measure of the modularity is then
\begin{equation}
Q=\Sigma_i(e_{ii} - \alpha_i^2),
\label{modu}
\end{equation}
which calculates the fraction of edges which connect nodes of the same type minus
the same quantity in a network with  identical  community division
but random connections between the nodes. If the number of within-community edges is no better
than random, $Q=0$. 
The modularity plotted at every level of the bifurcation would
indicate the quality of the detection procedure and at which level
to look at to see the best division where $Q$ has a maximum value.

There are some typical networks to which the community detection 
algorithms are applied to see how efficient it is. Most of 
the methods are applied to an artificial   
computer generated graph  to check its efficiency. 
Here  128 nodes    are divided into    four  communities. Each node
has $k=16$ neighbours of which $k_{out}$ connections
are made to nodes belonging to other communities. 
 The quality of 
separation into communities here can be measured by calculating the
number of  correctly classified nodes. All available detection algorithms work very well for 
 $k_{out}\leq 6 $
and the performance becomes poorer  for larger values.
A comparative study of application of different algorithms to this 
network has been made by Danon et al \cite{danon}.

A popular network data with known community structure which is 
also very much in use to check or compare algorithms  is the 
Zachary's karate club (ZKC) data.
Over the course of two years in the early 70's, Wayne Zachary observed social interactions between
the members of a karate club at an American university \cite{Zach}. He constructed  network
of ties between members of the club based on their
social interactions both within and outside the club. As a result of a dispute between 
the administrator and the chief karate teacher the club eventually split into two. 
Interestingly, not just 
two, but upto five communities have been detected in some of the  algorithms.

Community structure in jazz network has also been studied. 
In this network,  musicians are the nodes and two musicians are connected if they have performed in the same band. 
In a  different formulation, the bands act as nodes and 
have a link if they have at least one  common musician.
This was studied mainly in \cite{jazz}.

The American college football network is also a standard network where 
community detection algorithms have been applied. 
This is a network representing the schedule of games 
between American college football teams in a single season.
The teams are divided into groups or ``conferences'' and
intra-conference groups are more frequent than inter-conference games. The
conferences are the communities, so that teams belonging to the same
conference would play more between themselves.

Community detection algorithms
have been applied to other familiar social networks like collaboration 
and e-mail networks as well as in several non-social networks.

\subsection{Some novel community detection algorithms}

The method of GN to detect the communities typically
involves a time of the order of $N^2M$, where $N$ is the number of nodes and $M$
the number of edges.  This is considerably  slow. Several other
methods have been proposed although none of them is really `fast'.
Some of these methods are just variants of the divisive
method of GN (i.e., edge removal by a certain rule)
and interested readers will find the details in  references \cite{NG,rad,fortunato}.
 Here we highlight those which are radically different from the GN method.

\noindent {\it {Optimisation methods}}

Newman \cite{newmanfast} suggested an alternative algorithm in which one attempts to 
optimise the splitting for which $Q$ is maximum.
However, this is a costly programme as the number of ways in which $N$ vertices
can be divided into $g$ non-empty groups is given by Stirling's number
of the second kind $S_N^{(g)}$ and the number of distinct
community division is $\Sigma_{g=1}^N S_N^{(g)}$.
It increases 
at least exponentially in $N$ and therefore approximate methods are required.
In \cite{newmanfast}, a greedy agglomerative algorithm was used. Starting from the  
state where each node belongs to its own cluster, clusters are joined 
such that the particular joining results in the greatest increase in $Q$.
In this way a dendogram is obtained and the corresponding $Q$ values are calculated. 
Community division for which $Q$ is maximum is then noted.
The computational effort involved in this method is $O((M+N)N)$. For small networks, the 
GN algorithm is better but for large networks this optimisation method functions more
effectively. Applications of this method has been done for the ZKC 
 and the American
football club networks.

Duch and Arenas \cite{Duch} have used another optimisation method which is local
using extremal optimisation. In this local method, a quantity $q_{r}$ for the $r$th node is defined
\begin{equation}
q_{r} = \kappa_{i(r)} - k_r\alpha_{i(r)},
\end{equation}
where $\kappa_{i(r)}$ is the number of links that a node $r$ belonging to 
a community $i$ has with nodes in the same community and  $k_r$ is the degree of
node $r$, $\alpha_{i(r)}$ is the fraction of edges connected to $r$ 
having either or both end in the community $i$. 
Note that $q_r$ is just the contribution to $Q$ (eq. (\ref{modu})) 
from each node  in the network,
given a 
certain partition into communities, 
and $Q=\frac{1}{2N} 
\Sigma _{r} q_{r}$. 
Rescaling the local variables $q_r$
by  $k_r$, a proper definition 
for the contribution of node $r$ to the modularity can be obtained. 
In this optimisation method, the whole graph is initially
split into two randomly. At every time step, the node with the lowest fitness 
$\lambda_r = q_r/k_r$ is moved from one partition to another. The process is repeated until an optimal 
state  with a maximum value of $Q$ is reached. After that all the 
inter-partition links are deleted and the  optimisation
procedure applied to the resulting partitions. The process is continued till
the modularity $Q$ cannot be improved anymore.
Applications were made to  the ZKC, jazz,  e-mail, cond-mat and biological
 networks. In each case  
it gives a higher modularity value compared to that of the GN method.

\noindent {\it {Spectral methods}}

The topology of a network with $N$ vertices can be 
expressed through a symmetric $N\times N$ matrix $L$, the Laplacian matrix, where  $L_{ii}=k_i$ (degree of the $i$th node) and  
 $L_{ij}$ are equal to $-1$ if $i$ and $j$ are connected and zero otherwise.
The column or row sum of $L$ is  trivially zero.
A constant vector, with all its elements identical will thus be
an eigenvector of this matrix with eigenvalue zero. 

\begin{figure}[b]
\includegraphics[clip, width=7cm]{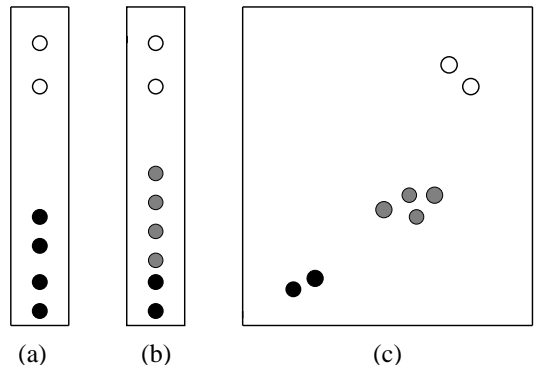}
\caption{(a) The structure of the first non-trivial eigenvector when there are
only two communities.   (b) An example of a case with three communities,
now the first nontrivial eigenvector has a structure with fuzziness when the components corresponding to 
two communities are pretty close (shown in gray and black). (c) When the
first two non-trivial eigenvectors are projected on a two dimensional space, the 
three communities are clearly shown.} 
\label{fig3}
\end{figure}
For a connected graph, the eigenvectors are non-degenerate. Since $L$ is
 real-symmetric, the eigenvectors are orthogonal. This immediately
implies that the sum of the components of the other eigenvectors are zero as
the first one is a constant one.  In case  the number of subgraphs is equal to 
two, there will be a clear cut bifurcation in the eigenvector; 
components of the second (first non-trivial) eigenvector are positive 
for one subgraph and negative for the other. 
This is the spectral bisection method which works very well when only two communities
are present, e.g., in the case of the ZKC network.

For more than two subgraphs, 
the distinction may become fuzzy (see Fig. \ref{fig3}). In this case, a few methods have been 
devised recently \cite{capocci,donetti1,donetti2}.
The eigenvectors of $L$ with non-zero eigenvalues 
are the non-trivial eigenvectors. If there is a network structure 
with $m$ communities, there will be $m-1$ nontrivial eigenvectors with 
eigenvalues  close to zero and these will have a characteristic structure:
the components corresponding to nodes within the
same cluster have very similar values $x_i$.  
If the partition is sharp, the profile of each eigenvector, 
sorted by components, is step like.
The number of steps in the profile corresponds to the number of communities.
In case the partition is not so sharp, and the number of nodes is high, it is
difficult to arrive at a concrete conclusion by just looking at the
first nontrivial eigenvector.
The method is then to combine information from the first few, say $D$, nontrivial eigenvectors.
Consider 
 an enlarged $D$  dimensional space where the components of these eigenvector are 
projected (i.e., the coordinates of the points in this space  
correspond to the values of the components  of the eigenvectors for   each node).
The points one gets  can now be treated  as the ``positions''
of the nodes and  nodes belonging to the same community are clustered
in this space.
Donetti and Munoz \cite{donetti1} 
used  an agglomerative method to detect communities  from this $D$ dimensional space.
The choice of $D$, the number of eigenvectors to be used, is an
important question here. One can fix a value of $D$, and find out the optimal
value of $Q$ - the $D$ value corresponding to the best optimal value 
of $Q$ may be the ideal choice.
It was 
observed  that $D$ is  of the order of the number of communities in the
network. 

Instead of the Laplacian matrix, one can also take another matrix, $X = K-A$
where $K$ is the diagonal matrix with elements 
$k_{ii} = \Sigma _{j=1}^N a_{ij}$ and $a_{ij}$ the elements of the adjacency
matrix $A$.   Here instead of
zero eigenvalue, the normal matrix $X$ has largest eigenvalue 1   associated with
the trivial eigenvector. 
This matrix has been used by Capocci et al \cite{capocci} for networks with weighted edges
successfully. It has been tested on a directed network of word association where the
method has been modified slightly to take into account the directedness of the edges.

\noindent {\it {  Methods based on dissimilarity}}

Two vertices are said to be similar if they have the same set of 
neighbours other than each other.
Exact similarity is rare, so one can define some dissimilarity
measure and find out the communities using this measure. 
One of these measures is called the Euclidean distance discussed 
in \cite{newman_review}. Recently, another measure based on the 
differences in their ``perspective'' was done.   Taking any vertex ``$i$'', the distance $d_{ik}$  to
a vertex ``k'' can be measured easily.
Taking another node ``$j$'', 
one can measure $d_{jk}$. If the nodes belong to the
same community, then the difference $[d_{ik} -d_{jk}]^2$ is 
expected to be small. The dissimilarity index $\lambda_{ij}$ was defined by 
Zhou \cite{Zhou}  as
\begin{equation}
\lambda_{ij}=\frac{\sqrt{\Sigma_k [d_{ik} -d_{jk}]^2}}{N-2}.
\end{equation} 
The algorithm is a divisive one involving   a few  intricate steps
and has been successfully applied to many networks.

\noindent {\it{Another  local method}}

Bagrow and Bollt \cite{bagrow}  
have suggested a local method in which, starting from a particular node, one finds out $k^{(l)}$, the number 
of its $l$th nearest neighbours  at every stage $l=0,1,2,...~ $.  By definition,
$k^{(0)} = 1$. The  cumulative number of degrees $K^{(l)} = \Sigma k^{(l)}$ 
is also  evaluated here.
The ratio $K^{(l+1)}/K^{(l)}$ is compared to a preassigned 
quantity $\alpha$ and in  case the ratio is less than $\alpha$,
the counting is terminated and all nodes upto the $l$th neighbourhood are listed as members of the
starting node's community.

\subsection{Community detection methods based on physics}

A few methods of detecting communities are based on models and methods used in 
Physics. 

\noindent {\it Network as an electric circuit}

Wu and Huberman \cite{WuHuber} 
have proposed a method by considering the graph to be an electric circuit. 
Suppose it   is known that nodes A and B belong to different communities.
Each edge of the the graph is regarded as a resistor and all edges  have identical  resistance.
Let a battery connect A and B such that the voltages at A and B are 1 and 0 respectively.
Now the graph is simply an electric circuit with a current flowing through
each resistor. By solving Kirchoff's equations one can 
obtain the voltage at each node. Choosing a threshold value of the voltage (which
obviously lies between 0 and 1), one then 
decides whether a  node belongs to the community containing A or that containing B.
It is easy to see how this method works: first suppose that there is a node C which is a leaf or peripheral node, i.e., connected to a single node D.
No current flows through CD and therefore C and D have the same voltage
and  belong to  the same community. Next consider the case when
C is connected to nodes D and E. Since the resistance is same for all edges,
$V_c = (V_D + V_E)/2$. Hence if both D and E belong to the same community 
(i.e., they both have high or low voltages), C also has a comparable voltage and
belongs to the same community.
If, however, $V_D$ and $V_E$ are quite different and D and E belong 
to different communities then it is hard to tell  which
community C belongs to. This is close to reality, a node has connection 
with more than one community.
In general, when C is linked to $n$ other nodes with voltages $V_1, V_2,...,V_n$,
$V_c = \Sigma V_i/n$. Hence if most of its neighbours belong to a particular community, C also belongs to it.
The method works in linear time.

When it is not known initially that two nodes belong to two different communities,
one can randomly choose two nodes and repeat the process of
evaluation of communities a large number of times. Chances that the detection is correct is fifty percent.
However, this probability will be greater than half if these two initial
nodes are chosen such that they are not connected. Majority of the results would then be regarded as correct.
Applications have been made to the ZKC  and the College football
club networks.

\noindent {\it Application of Potts and Ising models:}

The $q$ state Potts model in Physics is a model in which the  spin at  site
$i$ can have a value $\sigma_i = 1,2,....,q$. 
If two spins interact in such a way that the energy is minimum when they have the same value, we have
the ferromagnetic Potts model. Potts model has well known application 
for graph colouring problem, where each colour corresponds to a value of the spin. For community
detection also, the method can be similarly applied. 
Reichardt and Bornholdt \cite{reich} have used the Potts model Hamiltonian 
\begin{equation}
H = -J \Sigma_{i,j} \delta_{\sigma_1,\sigma_2} + \gamma \Sigma_\sigma ^q  
\frac{n_\sigma (n_\sigma-1)}{2},
\end{equation}
where $n_\sigma$ is the number of spins having spin $s$.
Here the first term is the standard ferromagnetic Potts model and
in absence of the second term it favours a homogeneous distribution of spins 
which minimises the ferromagnetic term. However, one needs to have diversity and that is incorporated  by the second term.
If there is only one community it will be maximum and if spin values 
 are more or 
less evenly distributed over all nodes it will be  minimum.

When $\gamma$ is set equal to the average connection probability, then 
both the criteria  that (a)  within a community the connectivity is larger than
the average connectivity, and that (b) the   average connectivity  is greater than the
 inter-community connectivity, are satisfied. 
With this value of $\gamma$ and $J=1$, the ground state was found out by the
method of  simulated annealing. 
The communities appeared as domains of equal spin value
near the ground state of the system. 
Applications
to computer generated graphs and biological networks were made successfully
and the method can work faster than many other algorithms.

In another study by Guimer\`a et al \cite{guimera} in which the Potts model 
Hamiltonian is used again,  although  in a 
slightly different form, an interesting result was obtained.   
It was shown that networks embedded in low-dimensional spaces have
high modularity and that even random graphs can have high modularity
due to fluctuations in the establishment of links. The authors
argue that like the clustering coefficient, 
 modularity in complex networks should also be compared to that of the  
 random graph.

Son et al \cite{son} have used a Ising model Hamiltonian to 
detect the community structure in networks.   The Ising model
is the case when the spins have only two states, 
usually taken to be $\sigma = \pm 1$. 
This method can be 
best understood in the context of the Zachary karate club case. 
Here two members started
having difference and the club members eventually got
polarised. In this case, the authors argue, the members will try to minimise 
the number of broken ties, or in other words, the breakup of ties should be 
in accordance to the community structure. Thus the community structure may be found out by simulating the breakup caused by an enforced frustration among nodes.
The breakup is simulated by studying the ferromagnetic random field Ising model
which has the Hamiltonian
\begin{equation}
H = -\frac{1}{2} \Sigma_{i,j} J_{ij} \sigma_i \sigma_j - \Sigma_i B_i \sigma_i,
\end{equation}
where $\sigma_i$ is the Ising spin variable for the $i$th node and the spins
interact ferromagnetically with strength $J_{ij}$ and $B$ is the random field. 
$J_{ij}$ is simply equal to the elements of the adjacency matrix, can be 1 or 0 for  unweighted networks and equal to the weights for a weighted graph.
The ferromagnetic interaction, as in the Potts case, represents the cost
for broken bonds. The conflict in the network (e.g., as raised in the 
ZKC) is mimicked in the system by imposing very strong negative
and positive fields for two nodes (in the ZKC case, these nodes should
correspond to the instructor and administrator).
The spins with the same signs in the ground state then belong to the 
same community.
The method was  applied to the Zachary Karate club network  
and the co-authorship networks with a fair amount of success.
It indicated the presence of some nodes which were marginal in the sense 
that they did not belong to a community uniquely.

\subsection{Overlap of communities and a network at a higher level}

\begin{center}
\begin{figure}
\noindent \includegraphics[clip,width= 7cm]{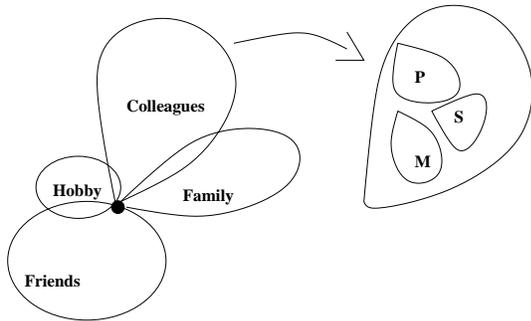}
\caption{A node (black circle) belonging to different communities. i
A particular community, e.g., the community of colleagues, 
may again contain subcommunities P,M,S etc. }
\label{fig4}
\end{figure}
\end{center}

Division of a society into communities may be done in many ways.
If the criteria is friendship, it is one, 
while if it is professional relation it is another. A node in general 
belongs  to more than one community. 
If one focuses on an individual node,  it lies at the center of 
different communities (see Fig. \ref{fig4}). Within each of these communities also,
there are subgraphs which may or may not have overlaps. 
Considering such a scenario, Pall\`a et al \cite{palla1} developed 
a new concept, the  ``network of communities'',   in which the communities
acted as nodes and the overlap 
between communities constituted the links.

Here a community was assumed to consist of several complete subgraphs 
($k$ cliques \cite{clique}) which tend to share many 
of their nodes and a single node belongs to more than one community.
Pall\`a et al \cite{palla1}  defined certain quantities:\\  
(i) The membership number $m$ which is the number of communities a node belongs to\\
(ii) Size of a community $n_s$ which is the number of nodes in it\\
(iii) Overlap size $s^{\alpha,\beta}$, the number of nodes shared by 
communities $\alpha $ and $\beta$.\\
Two quantities related  to the network of communities were also defined:\\
(i) Community degree $k_c$ - the number of communities a community is attached to.\\
(ii) Clustering coefficient of the community network.\\

A   method was used in which the $k$ cliques 
are first located and then the identification is carried 
out by a standard analysis of the clique-clique overlap matrix \cite{clique}.
Identification and the calculation of the distributions for the quantities 
$m$, $n_s$, $s^{\alpha,\beta}$ and $k_c$ were done for the co-authorship, 
word association  and  biological networks. 

\vskip -0.5cm

The community size distributions $P(n_s)$ for the different real networks 
appear to be power law distributed ($n_s^{-\tau}$)   with the value of the exponent $\tau$ between $-2$ and $-2.6$. 
At the node level, the degree distribution is often a power
law in many networks. Whether at a higher level, i.e., at the 
community level it also has the same organisation is the important question. 
The degree distribution of the communities showed  
an initial  exponential decay  
followed by a power law behaviour with the same exponent $\tau$.

The distribution of the overlap size $s^{ov}$ and the membership 
number $m$ also  
showed power law variations with both the distributions having a large exponent
for the collaboration and word association networks. This shows that
there is no characteristic overlap size and that a node
can be the member of a large number of communities. 

The clustering coefficients of the community networks were found out to be  fairly large, e.g., 
0.44 for the  cond-mat  collaboration network, showing that 
two communities overlapping with a given community has a large probability
of overlapping with each other as well.  

\vskip 0.5cm

\noindent  {\it Preferential attachment of communities }

As the community degree distribution (or the size distribution) 
is scale-free, the next question which arises is what happens to
a new node which joins a community, i.e., what is the attachment rule chosen for the new node as 
regards the size and degree of the communities \cite{palla2}?

To determine the attachment probability with respect to
any property $\rho$, the cumulative distributions $P(\rho)$ at times $t$ and $t+1$ are noted. 
Let the  un-normalised $\rho$ distribution of the chosen objects during the
time interval $t$ and $t+1$ be $w_{t\to t+1}(\rho)$.
The value of $w_{t\to t+1}(\rho^*)$ at a given $\rho^*$
equals the number of objects chosen which have a $\rho$ value greater
than $\rho^*$.
If the process is uniform, then objects chosen with a given $\rho$ 
are chosen at a rate given by the distribution of $\rho$ amongst the available
objects. However, if the attachment mechanism 
prefers high (low) $\rho$ values, then objects with high (low) $\rho$ 
are chosen with a higher rate compared to the distribution 
$P(\rho )$.
In real systems, the application of the method indicated 
that rather than a uniform attachment
  there is a preferential
attachment even at the higher level of networks.

\section{Models of social network}

In this section we discuss some models of social network which particularly
aim at reproducing the community structure and/or positive assortativity.
We also review some  dynamical models which show rich phase transitions.

\subsection{Static models}
\label{static}

By static models we mean a social model where 
the existing relations between people  remain unchanged while new 
members can enter and form links and the size of the system may grow.
Examples are collaboration networks where existing links cannot be rewired.

Newman and Park \cite{newman_park} have constructed a model  which has  
a given community division.
The aim is to show that it has positive assortativity.
Here it is assumed that members belonging to the same community are 
linked with probability $p$ similar to 
a bond percolation \cite{perco} problem. 
An individual may be attached to more than one
community, 
 and the  number of members in a community is
assumed to be a variable here. 

The assortativity coefficient $r$ was calculated here in terms of $p$ and 
the moments 
of the distributions of  
$m$ (the number of communities to which a member belongs) and  
 $s$ (the community size). 
The theoretical formula was applied to real systems after  estimating the  distributions 
by  
detecting communities  using standard algorithms.
The value of $p$
was calculated by dividing the number of edges in the network by the
total number of possible within-group edges. 
The theoretical value of $r$ for the co-authorship network turns out to be
0.145 
which is within the statistical error of the real value 
$0.174 \pm 0.045$.

Another model was proposed by Catanzaro et al \cite{Catan}
to reproduce the observations of a specific database, the cond-mat archive 
of preprints. This network of co-authors was found to be scale-free
with positive assortativity. In the model the 
preferential attachment scheme was used with some modification: the new node 
gets linked  to an existing node with preferential
attachment but  with a probability $p$.  Two existing nodes with degree $k_1$ and $k_2$  could also 
get linked 
stochastically with the probability   being proportional  
to  $(1-p)f(|k_1-k_2|)$.
$f(x)$ was chosen to be decaying with   $x$ either exponentially or as a inverse power law.
It was shown that by tuning the parameter $p$, one could obtain different
values of the exponents for the degree distribution, clustering coefficient,
assortativity and betweenness.  The results agree fairly well for all the 
quantities of the actual network except the clustering coefficient.

Bog\~una et al \cite{Boguna} have used the concept of social distance in their model. 
A set of quantities $h^i_n$ for the $n$th individual is used to represent 
the characteristic features of individuals like profession, religion, geographic locations etc. 
Social distance between two individuals $n$ and $m$ 
with respect to characteristic $i$ can be quantified by
the difference of $h_n^i$ and $h_m^i$. It is assumed that 
two individuals with larger social distance will have a lesser probability
of getting acquainted. If the social distance with respect to the $i$th
feature is denoted by $d_i(h_n^i, h_m^i)$, then a connection probability is defined
as 
\begin{equation}
r_i(h_n^i, h_m^i) = \frac{1}{1+[b_i^{-1}d_i(h_n^i, h_m^i)]^\alpha},
\end{equation} 
where $\alpha$ is a parameter measuring homophily,  the tendency of
people to connect to similar people and $b_i^{-1} $ is a characteristic 
scale.
The total probability of a link to exist between the two individuals is a weighted sum
of these individual probabilities
\begin{equation}
r(h_n, h_m) = \Sigma_i w_i r_i(h_n^i, h_m^i).
\end{equation}
The degree distribution, the clustering coefficient and the assortativity (in terms of $\langle k_{nn}\rangle$) 
were obtained analytically
for the model and also compared to simulation results. The resulting 
degree distribution had a cutoff, the   clustering coefficient 
showed increase with $\alpha$ and the assortativity was found out to be positive. 
The model also displayed a community structure when tested with 
the GN algorithm.

In the model proposed by Wong  et al \cite{ling}, nodes are distributed in a Euclidean space and the 
connection probabilities depend on the distance separating them. Typically,
a neighbourhood radius is defined within which nodes
are connected with a higher probability. 
For chosen values of the parameters in the scheme, 
graphs were generated which showed community structure and 
small world effect.

\begin{figure}
\noindent \includegraphics[clip,width= 8cm]{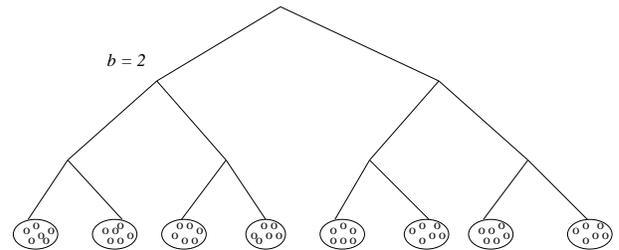}
\caption{A hierarchical structure of human population obtained using 
a single characteristic feature. People belonging to the same lowest
group (ellipse) have social distance 1 and the distance increases as one goes up in the hierarchy.
The maximum distance in this case is $l=4$.}
\label{fig5}
\end{figure}

A hierarchical structure of the society has been assumed by Motter et al \cite{motter} in their model 
in which the concept of social characteristics is used. Here a community of $N$ people are assumed to have 
$H$ relevant social characteristics. Each of these characteristic defines a nested hierarchical organisation of groups, 
where people are split into smaller and smaller subgroups downwards 
in this nested structure (Fig. \ref{fig5}).
Such a hierarchy is characterised by the number $l$ of levels, the branching ratio $b$ at each level
and the average number $g$ of people in the lowest group.

Here also distance factors corresponding to each feature can be defined and it is assumed that the distances are 
correlated, i.e., two individuals are expected 
to be different (similar) to each other in the same degree with respect to 
different characteristics. 
Connection probability decreases exponentially with the social distance 
and links are generated until the number of links per
person reaches a preassigned average value. Networks belonging to the
random and regular class are obtained for extreme values of the parameters
and in between there is a wide region  in which  social 
networks fall.

A growing model of collaboration network to simulate the 
time dependent feature of the link length distribution observed in \cite{anjan}
has been proposed recently by Chandra et al \cite{anjan-model}. 
Here the nodes are embedded on a Euclidean two dimensional space
and an incoming node $i$ gets attached to (a) its nearest neighbour $j$,
(b) to the neighbours of $j$ with a certain constant probability 
and (c) to other
nodes with a time dependent probability. Step (a) takes care of 
the proximity bias, step (b) ensures high clustering 
and step (c) takes care of the fact that long distance collaborations increase
in time thanks to rapid progress in communication.
The link length distributions obtained from the simulations
are consistent with the observed results and the model also gives a
positive assortativity.

\subsection{Dynamical models}

Dynamical models allow the links to change in time 
whether or not the network remains constant in size.
This is close to reality as human interactions are by no means static \cite{holme,roth}.
These models are applicable to cases where direct human interactions 
are involved.

In Jin et al \cite{jinetal}, the model consists of a fixed number of members  
and  links may have a finite lifetime. Here it is assumed that the probability 
of interaction between two individuals
depends on their degrees and is enhanced by the  presence of a mutual friend.
A cutoff in the number of acquaintances ensures that the total number of links has
an upper bound. The strength $s_{ij}$ of a tie between individuals $i$ and $j$ 
is a function of the time $t$ since they last met; $s_{ij} \sim \exp(-\kappa t)$.
Starting with a set of individuals with no links, edges are allowed to appear
with a small value of $\kappa$ close to zero. When the network saturates, $\kappa$ is set
to a larger value and the evolution of the network studied.
The resulting network   shows features of large clustering and community structure 
for realistic  range of values of $\kappa$.

A somewhat similar model has been proposed by Ebel et al \cite{ebel-dyn} where new acquaintances 
are initiated by mutual friends with the additional assumption that
a member may leave the network with probability $p$. All its links are deleted in that case.
The finite lifetime of links brings the network to a stationary state which manifests large clustering 
coefficients, small diameter and scale free or exponential degree
distribution  depending on the value of $p$.

Another model which aims at forming communities is due to Gr\"onlund and Holme \cite{seceder} 
who construct a model based on the assumption that
individual people's psychology is to  try to be different from the crowd.
This model is based on the agent based model known as the seceder 
model \cite{seceder-original}. 
It involves an iterative scheme where  the  number of nodes remains constant and the links are 
generated and rewired during the evolution.
Networks generated by this algorithm  indeed
showed community structure and  small world effect while the degree distribution
had an apparent exponential cutoff. The clustering coefficient was found to be 
much higher than the corresponding random network while the assortativity 
coefficient $r$ showed a positive value.

Some of the dynamic models are motivated by the model proposed by Bonabeau et al \cite{Bona} in which agents are distributed randomly
in a $L \times L$ lattice and assigned a ``fitness'' variable $h$. The agents perform
random walk on the lattice and during an interaction of agents $i$ and $j$, the probability that 
$i$ wins over $j$ is 
$q_i$ given by  
\begin{equation}
q_i = \frac{1}{1+\exp(\eta (h_i -h_j))}.
\end{equation}
If the fluctuation $\sigma$ of  the $q_i$ values is small, the resulting society is egalitarian while for
large values of $q$, it is a society with hierarchical structure. In the model  
the individuals occupy the lattice with probability $p$. A phase transition is
observed as the value of $p$ is changed  (in a slightly modified model where $\eta$ is replaced
by $\sigma$). Gallos \cite{gallos} has shown that 
when a model with random
 long distant connections
is considered 
the phase transition remains 
with subtle differences in the behaviour of the distribution of $q$. 

In another variation of the Bonabeau model, Ben-Naim and Redner \cite{redner3} allow the fitness variable 
to increase by interactions and 
decline  by inactivity. The latter occurs with   a  rate $r$. The rate equation for the fraction of 
agents with a given fitness was solved to show a phase transition as $r$ is varied; with 
$r \geq  1$, one has a society with a single class and for $r \geq  1$, a hierarchical multiple
class society can exist. In the latter case,
the lower class is destitute and the middle class is dynamic and has a continuous upward mobility.

Individuals are endowed with a complex set of characteristics rather than
a single  fitness variable.
Models which allow dynamics of  these social characteristics
have also been proposed recently. Transitions from a perfect homogeneous 
society to a heterogeneous society has been found in the works of 
Clemmy et al \cite{klem} and Gonzalez-Avella et al \cite{klem2}.

Let each trait  take any of the integral
values $1, 2,..., q$. 
When individuals have nearest neighbour interactions  and the initial traits follow
a Poisson distribution, 
 the non-uniformity in the traits can drive the system to
a  heterogeneous state at $q=q_c$ \cite{axel}.   
Klemm et al found that on a small world network, the transition point $q_c$ shifts towards higher values
as the  disorder (fraction
of long range bonds) is enhanced
and the  transition  disappears  in a scale-free BA 
network in the thermodynamic limit.
In \cite{klem2}, the additional influence of an external field has been considered.

Shafee \cite{shafee} has considered a spin glass like model of social
network which  is described by the Hamiltonian
\begin{equation}
H = - \Sigma_{i,k,a,b} J^{ab}_{ik} s^a_i s^b_k - \Sigma_{i,a} h_i^a s^a_i, 
\end{equation}
where  $s^a_i$ is the state of the $i$th agent with respect to trait $a$, 
$J^{ab}_{ik}$ are the interaction between different agents and $h_i^a$ an
external field. Zero temperature Monte Carlo dynamics is then applied to
the system. However, the simulations were restricted to systems of very small sizes.
Results showed either punctuated equilibrium or oscillatory behaviour  of the trait values with
time.

\section{Is it really a small world? Searching: post Milgram}

Although Milgram's experiments have been responsible for inspiring
 research in small world networks, they have their own limitation. 
The chain completion 
percentage was too low - in the first experiment it was just five.  The 
later experiment had  a little better success rate but this time the recruit
selection was by no means perfectly random. 
Actually, a number of subtle effects can have profound bearing on what can be termed ``searching in small worlds''.
First, although the network may have a small world property, 
searches are done mainly
locally - the individual may not know the global structure of the network
that would help her or him to find out the shortest path to
a target node. 
Secondly, many factors like the race, income, family connections,
job connections, friend circle etc. determine the dynamics of the search process. Quite a few experiments in this direction are being carried out currently.  
Some theoretical approaches have also been
 proposed for the searching mechanism.

\subsection{Searching in small world networks}

The first theoretical attempt to find out how a search procedure 
works in a small world network was made by Kleinberg \cite{klein} 
in which it was assumed that the nodes were embedded on a two dimensional 
lattice. Each node was connected to its nearest neighbour and 
a few short cuts (long range connections) were also present to
facilitate the small world effect. 
The probability that two
nodes at an Euclidean distance $l$ had a connection was 
taken to be $P(l) \sim l^{-\alpha}$. 
The algorithm used was a ``greedy algorithm'', i.e., 
each node would send  the message to one of its nearest neighbours, and
look for the neighbour which takes it  closest to the target node.
Here the source and target nodes are random.
Interestingly, it was found that only for a special value
of $\alpha=2$, the time taken (or the number of steps) scaled as $\log(N)$,
while for all other 
values it varied sublinearly with $N$.
This result, according to \cite{klein}
could be generalized to any lattice of dimension $d$ and
there always existed a unique value of $\alpha = d$ 
where short paths between two random nodes could be found out.
Navigation on 
the WS model \cite{moura}  
and a one-dimensional Euclidean network \cite{zhu}  
agreed perfectly with the 
above picture.

The above results indicate that although the network may have small
average shortest paths globally, it does not necessarily mean that 
short chains 
can be realised  using local information only. 

\subsection{Searching in scale-free graphs}

In a scale free network with degree distribution given by 
$P(k) \sim k^{-\tau}$, one can consider two kinds of searches, one random and 
the other one biased. In the latter, one 
has the option of choosing neighbouring nodes with higher degree which 
definitely makes  the searching process more efficient.

Adamic et al \cite{adamic1} have assumed that apart from the 
knowledge about one's nearest neighbours, 
each member also has
some knowledge about the contacts of the 
second neigbours. 
Using this assumption, 
the  average search length $s$ 
in the random case 
is given by
\begin{equation}
s \sim N^{3(1-2/\tau)}.
\end{equation}
Only for $\tau=2$, the search length is $O(\ln(N)^2)$. 
The corresponding expression for the biased search, 
when nodes with larger degree are used, is  
\begin{equation}
s \sim N^{2-4/\tau}.
\end{equation}
 Simulations were done with  random and biased  search mechanisms where 
a node could scan both its first and second neighbours. The results 
for $\tau \neq 2$ indeed  yielded shortest paths which scale sublinearly 
with $N$, although in a slower manner compared to the theoretical predictions. 

\subsection{Search in a social network}

A social network comprising of people whose links are based on friendship or
acquaintance can be thought of as a network where the typical degree of each
node is $k=O(10^3)$. Hence the number of second neighbours should be
of $k^2$ and therefore, ideally in two steps one can access $10^6$ number of 
other people and the search procedure would be complete if the order of
total population is similar. Even if the degree is one order lesser, the number of steps is still quite small and search paths can be further shortened by taking advantage of highly connected individuals.
However, there is a flaw in this argument as there is a 
considerable  overlap of the  set of one's friends and that of 
one's friends' friends. 

It is expected that the participants in a real searching procedure 
would like to take advantage of certain features like geographical proximity or
similarity of features like profession and hobbies etc. In this background the hierarchical structure of the social network is extremely significant (Fig. \ref{fig5}).

Watts et al \cite{watts-search} have considered a hierarchical  model in 
which the 
individuals are endowed with network ties and identities by a set of characteristics as
described in sec. \ref{static}.  Distance  between individuals are calculated as the height of their lowest common ancestor level in the hierarchy.

The probability of acquaintance between two individuals is assumed to be 
proportional to  $\exp(-\alpha x) $ where $\alpha$ is the measure of homophily
and $x$ the social distance. 
 Unlike \cite{motter},   here it is assumed that the 
distance between two individuals for two different features are uncorrelated.
One can consider the minimum distance $y_{ij}$, the smallest of  the 
$x_{ij}$'s corresponding  to all the social features, to be sufficient 
to connote affiliation. 

It is assumed that the individual who passes the message knows only its own coordinates, its neighours coordinates and 
the coordinates of the target node. 
Thus the search process is based on partial information, information about 
social distance and network paths are both only locally known.

In this work one important aspect of the original experiments by Milgram and 
his coworkers was brought under consideration, that most of the 
search attempts failed but when it was successful it took only a few steps.
Introducing a failure probability $p$ that a node fails to 
carry forward the message, it is to be noted that  if the probability of a successful 
search of length $s$ is $r$, it must satisfy $q = (1-p)^s \geq r$. 
This gives an upper limit for $s$; $s \leq \log(r)/\log(1-p)$. 
In the simulations, the number of traits $H$ and $\alpha$ were 
varied keeping $p=0.25$ and $r=0.05$ 
fixed (these values are in accordance with realistic values which 
gives $s \leq 10.4$). The values of the average number of nearest neighbours and
branching ratio were also kept constant. A phase diagram in the $H-\alpha$ plane showed regions where the 
searching procedure can be successful. It showed that almost all searchable networks display $\alpha > 0$ and $H > 1$. 
The best performance, over the largest range of $\alpha$, is achieved for $H=2$ or $3$.
In fact the model could be tuned to reproduce the experimental results 
of Milgram ($s \sim 6.7$).

\subsection{Experimental studies of searching}

 A few projects to study  searching in social networks experimentally have
been initiated recently. Dodds et al \cite{dodds} have conducted a global, Internet based social search by registering
participants online. Eighteen target persons from 13 countries 
with varied occupation were randomly selected and the participants were
informed that their task was to help relay a 
message to their allocated target by passing the message
to a social acquaintance whom they considered closest to the target. 
One fourth of the participants   provided their personal information and
it showed that the sample was sufficiently representative of the
general Internet using population.
The participants also provided data as to on what basis  he or she chose the 
contacts' name and e-mail address; in maximum cases they were friends, 
followed by relatives and
colleagues.

The links in this experimental network showed that geographical proximity
of the acquaintance to the target and similarity of occupation
were the two major  deciding factors behind their existence.
Many of the chains were terminated and not completed as in Milgram's experiment
and the reason behind it was mainly lack of interest or incentive. In total, 384 chains were completed (nearly 100000 persons registered in the beginning).
It was also found that when chains did complete, they were short, the average
length $s$ being 4.05. However, this is a measure for completed 
chains only, and the hypothetical  estimate in the limit of zero
attrition comes out to be $s = 7$.  

In another study, Adamic and Adar \cite{adamic-search} 
derived the social network of e-mails at HP laboratories from the e-mail logs by defining 
a social contact to be someone with whom
at least six e-mails have been exchanged both ways over an approximate period of three months.
A network of 430 individuals was generated and the degree distribution
showed an exponential tail.
Search experiments were simulated on this network. Three criteria for
sending messages in the  search 
strategy was tested in  this simulation; degree of the node,
closeness to the target in the organisational hierarchy and location with
respect to the target. 

In a scale-free network, seeking a high degree node was shown to be a 
good search strategy \cite{adamic1}. However, in this network, which has an exponential tail, this does not work out to be effective. This is because most of the nodes do not have a neighbour with high degree. The second strategy of utilising the organising hierarchy worked 
much better showing that the hierarchical structure in this network is 
quite appropriate. The probability of linking as a function of the separation
in the organisational hierarchy also showed consistency with an exponential 
decay as in \cite{watts-search}. The corresponding exponent $\alpha$ has a 
value = 0.94 and is well within the searchable region identified in \cite{watts-search}.

The relation between linkage probability and distance $r$ turned out to 
be $1/r$ in contrast to $1/r^2$ where the search strategy should work best according to \cite{klein}.
Using geographical proximity as a search strategy gave larger short paths 
compared to the search based on organisational hierarchy, but the path lengths
were still `short'.

A similar experiment was conducted  by Liben-Nowell et al \cite{liben}, 
using a real world friendship network to see how
the geographical routing alone is able to give rise to short paths. In this
simulated experiment, termination of chains was allowed. Chain
completion was successful in thirteen percent cases with average
search length a little below 6.

\section{Endnote}

Study of social networks has great practical implications with respect to
many phenomena like spreading of information, epidemic dynamics, 
behaviour
under attack etc. In these examples, the analysis of the corresponding
networks  can help in making  proper 
strategies for voting, vaccination  or defence programmes.
In this particular review,  we have 
highlighted mainly the theoretical aspects of social networks 
like structure, modelling, 
phase transitions and searching. 
Social networks has emerged  truly as an interdisciplinary 
topic during recent times and the present review is a little biased 
as it is from  the viewpoint of a physicist.

\section{Appendix: The Indian railways network}

\begin{figure}[b]
\includegraphics[width=5cm]{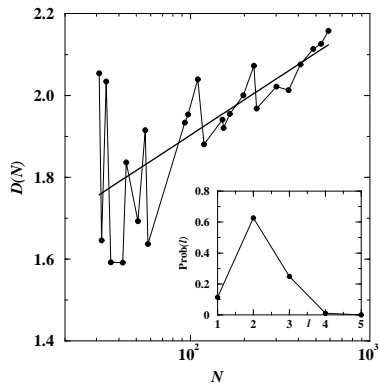}
\caption{
The variation of the mean distance ${\cal D}(N)$ of 25 different subsets of IRN having
different number of nodes $(N)$. The whole range is fitted with a function like
${\cal D}(N) = A+B\log(N)$ where $A \approx 1.33$ and $B \approx 0.13$.
The inset shows the distribution Prob$(\ell)$ of the shortest path lengths $\ell$
on IRN. The lengths varied to a maximum of only five link lengths
and the network has a mean distance ${\cal D}(N) \approx 2.16$.
}
\label{train1}
\end{figure}

The Indian railways network (IRN) is more than 150 years old and has a large 
number of stations  and trains (running at both short and long distances).
In the study of the IRN \cite{train}, a coarse graining was made by selecting 
a sample of trains, and the stations through which these run.
The total number of trains considered was 579 which run through 587 stations.
Here the stations represent the `nodes'
of the graph, whereas two arbitrary stations are considered to be connected
by a `link' when there is at least one train which stops at both the stations.
 These two stations are considered to be at unit distance of separation irrespective
   of the geographical distance between them. Therefore the shortest distance
   $\ell_{ij}$ between an arbitrary pair of
   stations $s_i$ and $s_j$ is the minimum number of different trains one needs to board
   on to travel from $s_i$ to $s_j$.
 Smaller subsets
of the network were also considered to analyse the behaviour of different quantities as a function
of the number of nodes.
The average  distance between an
   arbitrary pair of stations  was found to 
    depend only logarithmically on the total number of stations in the country
as shown in Fig. \ref{train1}.

Like other social networks, the IRN also showed a large clustering 
coefficient which also depends on the number of nodes (stations). 
For the entire IRN,  the clustering coefficient is around 0.69,
the value for the corresponding random graph being 0.11. 
The clustering coefficient as a function of the degree $k$ of a node
showed that it is a constant for  small $k$ and decreases 
logarithmically with $k$ for larger values.

\begin{figure}
\includegraphics[width=5cm]{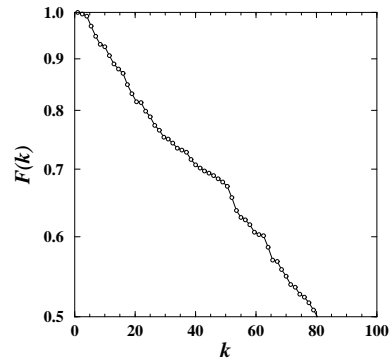}
\caption{
The cumulative degree distribution $F(k)$ of the IRN with the degree $k$ is
plotted on the semi-logarithmic scale.
}
\label{train2}
\end{figure}

The degree distribution $P(k)$ of the network, that is, the distribution of the
   number of stations $k$ which are connected by direct trains to an 
arbitrary station was also studied.
 The cumulative degree distribution
   $F(k) = \int_k^{\infty} P(k)dk $ for the whole IRN
  approximately fits to an exponentially decaying distribution:
   $F(k) \sim \exp(-\alpha k)$ with $\alpha$= 0.0085 (Fig. \ref{train2}).

\begin{figure}[b]
\includegraphics[width=5cm]{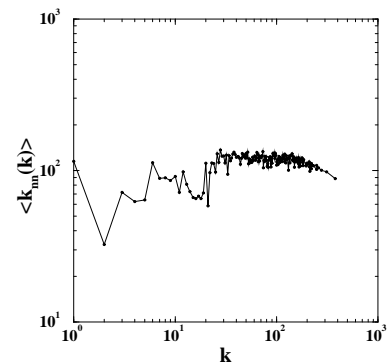}
\caption{
The variation of the average degree $\langle k_{nn}(k) \rangle$ of the neighbours of a node of
degree $k$ with $k$ in the IRN. After some initial fluctuations,  $\langle k_{nn}(k) \rangle$ remains
almost same over a decade around $k$ = 30 to 300 indicating absence of correlations
among the nodes of different degrees.
}
\label{train3}
\end{figure}

The average degree
   $\langle k_{nn}(k) \rangle$ of the nearest neighbours of a node
with degree $k$ is plotted in Fig. \ref{train3} 
to check the assortativity behaviour of the network. This
 data is not very indicative.  
The  assortativity  coefficient $r$ is therefore calculated 
using eq. (\ref{assort}), which gives the value $r$ = -0.033.
This shows that unlike social networks of class A and B, here 
the assortativity is negative.

\end{multicols}
\end{document}